\documentclass[twocolumn]{aastex63}
\usepackage{url}
\usepackage{hyperref}

\received{December 12, 2022} 
\submitjournal{ApJ}

\shorttitle{Circumnuclear region of the new high-redshift quasar
ULAS\,J0816+2134 at $z=7.46$}
\shortauthors{Koptelova \& Hwang}

\begin{document}


\title{Dense nitrogen-enriched circumnuclear region of the new
high-redshift quasar ULAS\,J0816+2134 at $z=7.46$}

\author{Ekaterina Koptelova}
\affiliation{Graduate Institute of Astronomy\\
National Central University \\
Taoyuan City 32001, Taiwan}


\author{Chorng-Yuan Hwang}
\affiliation{Graduate Institute of Astronomy\\
National Central University \\
Taoyuan City 32001, Taiwan}


\email{koptelova@astro.ncu.edu.tw}

\begin{abstract}
We present the $0.85-2.5$\,$\mu$m discovery spectrum and
multi-epoch photometry of the new high-redshift quasar
ULAS\,J081621.47+213442.6 obtained using the GNIRS spectrograph of
the Gemini North and near-infrared wide-field camera of the 4-m
UKIRT telescopes. The redshift of ULAS J081621.47+213442.6
measured from the Mg$\,${\scriptsize II}\,$\lambda2799$ emission
line is $z_{\rm Mg{\scriptsize II}}=7.461\pm0.007$. The absolute
magnitude of the quasar is $M_{1450}=-25.33\pm0.07$. The black
hole mass estimated using the Mg$\,${\scriptsize
II}\,$\lambda2799$ line and Eddington accretion rate are $M_{\rm
BH}\sim5\times10^{8}M_{\bigodot}$ and $L_{\rm bol}/L_{\rm
Edd}\sim0.7$. The spectrum of ULAS\,J081621.47+213442.6 exhibits
strong N$\,${\scriptsize III]}\,$\lambda1750$ emission line of a
rest-frame equivalent width of EW$_{0}\gtrsim12.5$\AA. The high
abundance of nitrogen suggests that ULAS\,J081621.47+213442.6 may
be at the peak of the nitrogen enrichment of the circumnuclear
region by the asymptotic giant branch stars, which is expected
$\sim0.25$\,Gyr after the bulk of star formation. The age of the
starburst of ULAS\,J081621.47+213442.6 implied by the high
nitrogen abundance, indicates that the active phase of the black
hole growth of the quasar may have lasted only $\sim0.25$\,Gyr,
favoring a massive initial black hole seed. We also observed the
flux variations of the UV continuum of ULAS\,J081621.47+213442.6
caused by the variation in the line-of-sight absorbing column
density on a rest-frame timescale of $\sim47$\,d. The estimated
hydrogen column density of the gas cloud responsible for this
variation is $N_{\rm H}\gtrsim10^{23.5}$\,cm$^{-2}$, consistent
with the typical column density of mostly neutral, gravitationally
bound clouds of the broad line region of quasars.
\end{abstract}


\section{Introduction}
\label{introduction}

High-redshift quasars at $z\lesssim7$ discovered in different sky
surveys
\citep[e.g.,][]{2010AJ....139..906W,2011Natur.474..616M,2013ApJ...779...24V,2015MNRAS.453.2259V,
2016ApJ...833..222J,2016ApJS..227...11B,2018ApJ...869L...9W,2019ApJ...883..183M,
2019ApJ...884...30W,2019AJ....157..236Y,2019MNRAS.487.1874R,
2022ApJS..259...18M} already harbor supermassive black holes
(SMBHs) of typical masses $10^{8}-10^{9}$$M_{\bigodot}$ at a
cosmic age of $\lesssim1$\,Gyr. The SMBHs of these quasars are
still in the phase of intensive growth as implied by their
near-Eddington accretion rates \citep[e.g.,
][]{2010AJ....140..546W,2011ApJ...739...56D,2017ApJ...849...91M}.
On the other hand, the discovery of the BL\,Lac object at
$z\approx6.6$ suggests that the SMBHs of some of the high-redshift
quasars may have already finished the growth by the end of the
reionization epoch \citep{2022ApJ...929L...7K}.

The massive SMBHs of $\sim 10^{9}$$M_{\bigodot}$ were also found
at the centers of the most distant of the known high-redshift
quasars, J1342+0928 at $z_{\rm [CII]}=7.541$
\citep[$M_{1450}=-26.76$; ][]{2018Natur.553..473B}, J1007+2115 at
$z_{\rm [CII]}=7.515$
\citep[$M_{1450}=-26.66$][]{2020ApJ...897L..14Y}, and J0313-1806
at $z_{\rm [CII]}=7.642$
\citep[$M_{1450}=-26.13$][]{2021ApJ...907L...1W}. The cosmic age
of $\lesssim$0.7\,Gyr corresponding to the redshifts of these
quasars provides strong constraints on the formation timescale of
the first SMBHs. However, the details of their formation process
remain elusive \citep[for a review, see ][]{2020ARAA..58...27I}.
Assuming the Eddington limited growth rate from $z\sim30$, the
$\sim 10^{9}$$M_{\bigodot}$ SMBH masses of these quasars should
have grown from initial seeds of $10^{3}-10^{5}$M$_{\bigodot}$
\citep[e.g.,][]{2021ApJ...907L...1W}.

The measurements of the metal abundance in high-redshift quasars
from different broad \citep[e.g.,
][]{2002ApJ...564..592H,2003A&A...398..891D,2003ApJ...589..722D,2006A&A...447..157N,2007AJ....134.1150J,2009A&A...494L..25J,2014ApJ...790..145D}
and narrow \citep[e.g.,
][]{2006A&A...447..863N,2009A&A...503..721M} emission lines are
consistent with the supersolar metallicity of their circumnuclear
gas even at $z\sim7.5$ \citep[e.g., ][]{2020ApJ...898..105O} and
imply the early chemical enrichment \citep[e.g.,
][]{1992ApJ...391L..53H,1993ApJ...418...11H}. For example, from
the eight line ratios with different nitrogen lines of 70
high-redshift quasars at $3.5<z<5.0$, \citet{2003ApJ...589..722D}
inferred a mean metallicity of the broad line region gas of
$5.3\pm0.3$Z$_{\bigodot}$ \citep[see also][]{2003A&A...398..891D}.
Similar metallicities were also derived for smaller samples, e.g.,
from the N$\,${\scriptsize V}/C$\,${\scriptsize IV} and
N$\,${\scriptsize V}/He$\,${\scriptsize II} line ratios of 6
quasars at $5.8<z<6.3$, \citet{2007AJ....134.1150J} inferred a
mean metallicity of $\sim$4Z$_{\bigodot}$, whereas
\citet{2009A&A...494L..25J} inferred a metallicity of
$\sim$7Z$_{\bigodot}$ from analysis of the (Si$\,${\scriptsize
IV}$\lambda$1397+O$\,${\scriptsize
IV]}$\lambda$1402)/C$\,${\scriptsize IV}$\lambda$1549 ratios of 30
quasars at $4.0<z<6.4$.

The method for constraining the epoch of first star formation in
the host galaxies of high-redshift quasars usually employs the
Fe$\,${\scriptsize II}/Mg$\,${\scriptsize II} line ratio and is
based on the delay in the Fe enrichment relative to Mg by
approximately 1\,Gyr \citep{1993ApJ...418...11H}. No evolution of
this line ratio with redshift was convincingly measured up to
$z\sim7.5$ \citep[e.g.,
][]{2003ApJ...596L.155M,2011ApJ...739...56D,2017ApJ...849...91M,2020ApJ...898..105O,2020ApJ...905...51S}.

Alternatively, the age of the starburst can be constrained based
on the abundance of nitrogen relative to C and O
\citep[][]{1978MNRAS.185P..77E,1992ApJ...391L..53H}. Nitrogen is
mainly synthesized as a secondary element from the previously
produced C and O, which results in the delayed nitrogen enrichment
with a characteristic time lag of $\sim0.25$\,Gyr
\citep{1991ApJ...367..126C,2000ApJ...541..660H,2022arXiv220204666J}.
The dominant source of nitrogen are stars in the asymptotic giant
branch (AGB) phase in a mass range of $4$M$_{\bigodot}\lesssim
m\lesssim 8$M$_{\bigodot}$ (with lifetimes $\sim2 \times 10^{8}$
and $\sim4 \times 10^{7}$\,yr, respectively)
\citep{1993ApJ...418...11H,1991ApJ...367..126C,2000ApJ...541..660H}.
These stars have higher metallicities compared to the preceding
generation of short-lived ($<4 \times 10^{7}$\,yr) massive stars
and therefore can produce larger amounts of N.

At the bulk of star formation, the chemical enrichment is driven
by massive stars with the largest C/N and the lowest N/O ratios.
The cosmological hydrodynamical simulations of chemical enrichment
predict the sharp local minimum in the C/N ratio at the onset of
the AGB stars when these stars start to release N
\citep{2018A&A...610L..16V,2018MNRAS.478..155V}. The chemical
evolution model of \citet{2009MNRAS.397.1661V} predicts that the
AGB stars may dominate the dust production on timescales of
$\gtrsim0.15$\,Gyr. After the onset of the AGB stars, the C/N
ratio is expected to increase due to the delayed release of C by
less massive AGB stars, and then decrease again on long timescales
due to the opposite dependence of the C and N stellar yields on
metallicity. Thus, the detection of the increased nitrogen
abundance expected $\sim0.25$\,Gyr after the bulk of star
formation can provide the approximate age of the starburst.

The observations show that the Eddington ratios of quasars
correlate with the abundance of nitrogen in their broad line
regions \citep{2011A&A...527A.100M,2017A&A...608A..90M}. Thus,
from the analysis of the optical spectra of the 2383 SDSS quasars
at $2.3 \lesssim z \lesssim 3.0$, \citet{2011A&A...527A.100M}
found that the quasars with the higher N$\,${\scriptsize
V}/C$\,${\scriptsize IV} and N$\,${\scriptsize
V}/He$\,${\scriptsize II} line ratios have higher Eddington
ratios. The black holes of some of the known high-redshift quasars
at $z\sim6.1-6.6$ showing high N$\,${\scriptsize
V}/C$\,${\scriptsize IV} line ratios are indeed among the most
actively growing SMBHs at early epochs accreting at the
super-Eddington accretion rates \citep[e.g.,
][]{2009ApJ...702..833K,2019ApJ...882..144K}.
\citet{2017A&A...608A..90M} also found that the Eddington ratios
and masses of the SMBHs of the nitrogen-rich quasars are
correspondingly high and low in comparison with normal quasars and
that the strong nitrogen lines of these quasars result not from
their high metallicities but from the high abundance of nitrogen
in their circumnuclear regions
\citep{2012A&A...543A.143A,2017A&A...608A..90M}. A similar
conclusion about the high nitrogen abundance of nitrogen-rich
quasars was obtained by \citet{2014MNRAS.439..771B}, although they
interpreted the high abundance of nitrogen as due to the generally
higher metallicities of nitrogen-rich quasars compared to normal
quasars. It was suggested that the high accretion rates of quasars
with the high relative abundance of nitrogen result from the mass
loss of the AGB stars
\citep{1993ApJ...416...26P,2007ApJ...671.1388D,2011A&A...527A.100M,2017A&A...608A..90M}.
The material released by these stars enriches the host galaxies of
quasars with nitrogen and also fuels the growth of their SMBHs.

The flux ratios between different broad emission lines depend not
only on metallicity, but also on physical conditions in the broad
line region such as the level of ionization, column density, and
number density of gas \citep[e.g., ][]{2020Atoms...8...94M}. The
broad line region gas is commonly represented as a system of line
emitting clouds of densities $n_{\rm
H}\sim10^{9}-10^{10}$\,cm$^{-3}$ and linear sizes of
$\sim10^{13}-10^{14}$\,cm orbiting at distances
$\sim10^{16}-10^{17}$\,cm from the central black hole \citep[for a
review, see, e.g.,][]{2008NewAR..52..257N,2009NewAR..53..140G}.
Observational evidence for the clumpy structure of the broad line
region is provided by the variation of the X-ray flux of Seyfert
galaxies caused by the motion of the absorbing/obscuring clouds
through the line of sight to the ionizing continuum
\citep{2002ApJ...571..234R,2007ApJ...659L.111R,2009ApJ...696..160R,2011MNRAS.410.1027R,2021ApJ...908L..33G}.
The observation of the partial obscuration of the X-ray continuum
emitting region by these clouds provides constraints on the
properties of the clouds. For instance,
\citet{2009ApJ...696..160R} showed that the variation of the X-ray
flux of the Seyfert galaxy NGC\,1365 was caused by the change in
the partial obscuration of the X-ray emitting region by a cloud of
size $\sim10^{13}$\,cm, column density of
$N_{H}\sim10^{23}$\,cm$^{-2}$, moving at a distance of
$\sim10^{16}$\,cm from the central black hole. The photoionization
calculations of \citet{2012MNRAS.424.2255W} predict that similar
to the X-ray emission, the UV emission of quasars can also be
significantly affected by absorption/obscuration caused by the
broad line region clouds. The observation of the obscuration of
the UV emitting region by such clouds can provide constraints on
the linear size and column density of the clouds similar to those
obtained, e.g., by
\citet{2007ApJ...659L.111R,2011MNRAS.410.1027R,2021ApJ...908L..33G}

In this work, we present the spectroscopic confirmation of a new
distant quasar, ULAS\,J081621.47+213442.6 at $z_{\rm
Mg\,{\scriptsize \rm II}}=7.461$ (hereafter ULAS\,J0816+2134). The
spectrum of ULAS\,J0816+2134 shows the strong N$\,${\scriptsize
III]}\,$\lambda$1750 emission line, which provides constraints on
the starburst age of its host galaxy. The multi-epoch observations
of the rest-frame UV continuum of ULAS\,J0816+2134 in the $J$
band, revealed significant variability attributed to the variation
in the line-of-sight absorbing column density. In
Section~\ref{sec:observations}, we describe the observational
data. In Section~\ref{sec:results}, we present the analysis of the
spectral properties and the $J$, $H$, and $K$-band light curves of
ULAS\,J0816+2134. In Section~\ref{sec:discussion}, we provide the
measurement of the black holes mass, discuss the growth history of
the black hole, and obtain constraints on the column density and
size of the gas clouds in the broad line region of
ULAS\,J0816+2134. In the paper, we assume a flat cosmology with
Hubble constant $H_{0}=70$\,km\,s$^{-1}$\,Mpc$^{-1}$, mass density
$\Omega_{\rm m}=0.3$, and vacuum density $\Omega_{\Lambda}=0.7$.

\section{OBSERVATIONS}
\label{sec:observations} \subsection{Broad-band Photometry}

ULAS J0816+2134 was selected using the catalogs of three different
surveys: the Panoramic Survey Telescope and Rapid Response System
1 \citep[PS1; ][]{2016arXiv161205560C}, United Kingdom Infrared
Telescope (UKIRT) Infrared Deep Sky Survey Large Area Survey
\citep[UKIDSS LAS; ][]{2007MNRAS.379.1599L}, and Wide-field
Infrared Survey Explorer \citep[WISE; ][]{2010AJ....140.1868W,
2014yCat.2328....0C}. The selection procedure employed the
following selection criteria: the detection in the UKIDSS $J$,
WISE $W1$, and $W2$ bands with approximate colors $J - W1 > -0.1$
and $W1 - W2
> 0$ AB mag expected for quasars at $z\gtrsim 6$, and the non-detection in all PS1 bands.
ULAS\,J0816+2134 was observed by UKIDSS LAS at two different
epochs, in 2007 in the $J$ band and in 2008 in the $Y$, $J$, $H$,
and $K$ bands. In 2007, the $J$-band brightness of this object was
below the detection limit of the survey. In 2008, ULAS\,J0816+2134
was detected in the $J$, $H$, and $K$ bands, but not in the $Y$
band (see Table~\ref{table:table1}). The non-detection in the PS1
and UKIDSS LAS $Y$ bands, infrared color, and $J$-band variability
of $>1$\,mag suggested that this object is a likely candidate for
a high-redshift quasar.

The deeper UKIRT $Z$-, $Y$-, $J$-, $H$-, and $K$-band images of
ULAS J0816+2134 were obtained on 2021 February 3, on 2022 January
27 and March 13. The near-infrared magnitudes of ULAS J0816+2134
from these observations are also presented in
Table~\ref{table:table1}. These recent images were reduced using
the same automated pipeline as the images of the UKIDSS LAS survey
\citep[see ][]{2006MNRAS.367..454H}. Note that the photometric
catalog released by the UKIDSS LAS survey provides only the
$J$-band magnitude of ULAS J0816+2134 corresponding to epoch 2008
(as given in Table~\ref{table:table1}). The $H$ and $K$ magnitudes
from the same epoch were not reported, likely due to the marginal
detection of ULAS J0816+2134 in both UKIDSS LAS $H$- and $K$-band
images. The UKIDSS LAS $H$ and $K$ magnitudes provided in
Table~\ref{table:table1} were estimated by us from the stack
images obtained by this sky survey and available at the UKIRT
WFCAM Science Archive (WSA)\footnote{http://wsa.roe.ac.uk/}. The
resulting magnitudes and the magnitude uncertainties quoted in the
table represent the average values and the standard deviations of
six different flux measurements relative to different nearby stars
in each of the $H$- and $K$-band UKIDSS images. The mid-infrared
magnitudes of ULAS\,J0816+2134 provided in
Table~\ref{table:table1} are from the AllWISE catalog of
\citet{2014yCat.2328....0C}.

\subsection{Near-infrared Spectroscopy}

The spectroscopic observations of ULAS\,J0816+2134  were conducted
on 2021 May 1, 4, and 17 using the Gemini Multi-Object
Spectrograph \citep[Gemini-N/GNIRS; ][]{2006SPIE.6269E..4CE} in
the cross-dispersed mode, which provided a spectral coverage of
$\sim0.85-2.5$ $\mu$m, and a slit width of 1$\arcsec$. The
resulting dataset consisted of $38$ spectroscopic exposures each
of $225$\,s obtained at a spectral resolution of
$R\thickapprox500$. The spectra were reduced using the Gemini IRAF
(Image Reduction and Analysis Facility) package. The telluric
correction was performed using two standard stars observed
together with ULAS\,J0816+2134 at a similar airmass: HIP\,35842
(type A1V, $J$ = 6.13 Vega mag) and HIP\,50459 (type A0V, $J$ =
6.54 Vega mag). The spectra obtained on different dates were
reduced and wavelength calibrated independently. The resulting
individual 2D spectra were matched in wavelength and stacked. The
final 1D spectrum and the error spectrum of ULAS\,J0816+2134 were
obtained from the 2D stack spectrum and its variance. The telluric
lines were removed by dividing the 1D spectrum of ULAS\,J0816+2134
by the 1D spectrum of HIP\,35842. The flux of ULAS\,J0816+2134 was
then recovered by multiplying the obtained ratio by the blackbody
model of the HIP\,50459 spectrum calculated assuming an effective
temperature of $T_{\rm eff}=9641$\,K reported for this star in the
Gaia DR2 catalog \citep{2018A&A...616A...1G}. To improve the
signal-to-noise ratio of the detected continuum, the final 1D
spectrum was binned to a wavelength resolution of 20\AA$ $
corresponding to full width at half-maxima of $\sim 480$, $\sim
370$, and $\sim 270$\,km\,s$^{-1}$ in the $J$, $H$, and $K$ bands.
The resulting spectrum is presented in Figure~\ref{figure:fig1}a.
The average signal-to-noise ratios of the spectrum are $\sim 3.5$,
$\sim 3.7$, and $\sim 6.9$ in the $J$, $H$, and $K$ bands,
respectively.

\begin{deluxetable}{ccccc}
\tablewidth{0pt} \tablecaption{Multi-epoch photometry of
ULAS\,J0816+2134.\label{table:table1}} \tablehead{
\colhead{Survey/} & \colhead{UT Obs. Date} &  \colhead{Filter} &
\colhead{$\lambda_{\rm eff}$} & \colhead{$m_{\rm AB}$} \\
\colhead{Telescope} & \colhead{(yy-mm-dd)} &  \colhead{} &
\colhead{($\mu$m)} & \colhead{(AB mag)} } \startdata
UKIRT   &  2021-02-03  &   $Z$&  0.88          &  $>22.78$ (5$\sigma$)      \\
UKIDSS  &  2008-11-28  &   $Y$& 1.03           &  $>21.10$ (5$\sigma$)      \\
UKIRT   &  $2021-2022$   &   $Y$& 1.03           &  21.683$\pm$0.121      \\
UKIDSS  &  2007-01-18  &   $J$& 1.25           &  $>21.15$ (5$\sigma$)     \\
UKIDSS  &  2008-11-28  &   $J$& 1.25           &  19.313$\pm$0.065     \\
UKIRT   &  2021-02-03  &   $J$& 1.25           &  21.427$\pm$0.035     \\
UKIRT   &  2022-01-27  &   $J$& 1.25           &  21.682$\pm$0.121     \\
UKIRT   &  2022-03-13  &   $J$& 1.25           &  21.314$\pm$0.068     \\
UKIDSS  &  2008-11-27  &   $H$& 1.63           &  21.085$\pm$0.123       \\
UKIRT   &  2021-02-03  &   $H$& 1.63           &  21.202$\pm$0.087       \\
UKIRT   &  2022-03-13  &   $H$& 1.63           &  21.093$\pm$0.126       \\
UKIDSS  &  2008-11-27  &   $K$& 2.20           &  20.573$\pm$0.095 \\
UKIRT   &  2021-02-03  &   $K$& 2.20           &  20.603$\pm$0.082 \\
UKIRT   &  2022-03-13  &   $K$& 2.20           &  20.611$\pm$0.114 \\
WISE          &  $2010-2011$   &   $W1$& 3.35 & 20.187$\pm$0.173      \\
WISE          &  $2010-2011$   &   $W2$& 4.60 & 20.036$\pm$0.312      \\
WISE          &  $2010-2011$   &   $W3$& 11.56 & $>17.49$ (2$\sigma$)      \\
WISE          &  $2010-2011$   &   $W4$& 22.09 & $>15.30$ (2$\sigma$)      \\
\enddata
\end{deluxetable}

\section{RESULTS}
\label{sec:results}

\subsection{Identification of the Emission Lines}
\label{sec:classification}

\begin{figure*}[ht] \vspace{-4.5cm} \hspace{-0.5cm}
\includegraphics[scale=2]{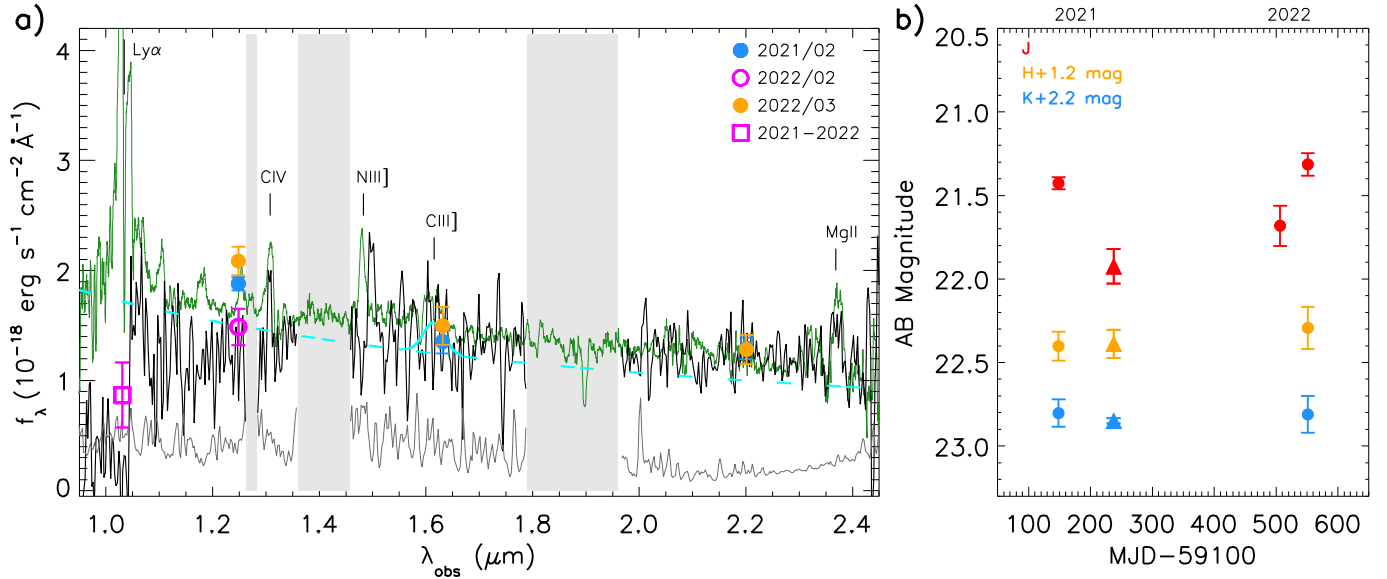}
\caption{(a) Flux-calibrated Gemini/GNIRS spectrum of ULAS
J0816+2134 (black line). The 1$\sigma$ error spectrum is shown
with a gray line. The wavelength intervals affected by strong
atmospheric absorption are shaded in light gray. The dashed cyan
line represents the $f_{\lambda}\propto \lambda^{-0.71}$ continuum
model. The Gaussian line shown in cyan on the top of the power-law
continuum is the Gaussian model of the C$\,${\scriptsize III]}
emission line of ULAS J0816+2134 calculated based on the predicted
flux and width of this line. The symbols show the $Y$-, $J$-,
$H$-, and $K$-band fluxes measured at different epochs between
2021 February and 2022 March. Overplotted in green is the scaled
and redshift-corrected SDSS spectrum of the low-redshift analog of
ULAS\,J0816+2134, the quasar SDSS J0852+4734 at $z_{\rm
Mg\,{\scriptsize II}}\approx2.428$ from the catalog of
nitrogen-rich quasars of \citet{2008ApJ...679..962J}. (b) The
$J$-, $H$-, and $K$-band light curves of ULAS\,J0816+2134 between
2021 February and 2022 March. The magnitudes measured from the
UKIRT images are shown with circles. Triangles correspond to the
$J$-, $H$-, and $K$-band magnitudes estimated from the
flux-calibrated Gemini/GNIRS spectrum. } \label{figure:fig1}
\end{figure*}

\begin{deluxetable}{lccc}
\tabletypesize{\footnotesize}
\tablecolumns{1} 
\tablewidth{\textwidth} \tablecaption{Redshifts, spectral slopes
($f_{\lambda}\propto \lambda^{\alpha_{\lambda}}$), and the
absolute magnitudes of ULAS\,J0816+2134 and comparison SDSS
quasars. \label{table:table2}}

\tablehead{\colhead{}                          &
           \colhead{$z_{\rm Mg{\scriptsize II}}$}                      &
           \colhead{$\alpha_{\lambda}$}                            &
           \colhead{$M_{\rm 1450}$}           }
\startdata
ULAS\,J0816+2134   &  7.461$\pm$0.007  & $-0.63$$\pm$0.44 &  $-25.33$$\pm$0.07  \\
\hline \multicolumn{4}{l}{Comparison SDSS quasars}\\
\hline
SDSS\,J0916+3908   &  1.876$\pm$0.001  & $-1.68$$\pm$0.11  &  $-24.94$$\pm$0.03  \\
SDSS\,J1550+4238   &  2.644$\pm$0.008  & $-0.97$$\pm$0.10  &  $-24.96$$\pm$0.03  \\
SDSS\,J0852+4734   &  2.428$\pm$0.001  & $-0.71$$\pm$0.15  &  $-25.06$$\pm$0.03  \\
\enddata
\end{deluxetable}

The spectrum of ULAS\,J0816+2134 shows the
Ly$\alpha$\,$\lambda$1216 break at $\sim10420$\,\AA $ $
corresponding to an approximate redshift of $\sim 7.57$ and three
emission lines. The emission lines at $\sim13068$ and 23698 \AA$ $
were identified as C$\,${\scriptsize IV}\,$\lambda$1549 and
Mg$\,${\scriptsize II}\,$\lambda$2799 both shifted to similar
redshifts of $\sim 7.44$ and $7.46$, respectively (see
Figures~\ref{figure:fig1}a and \ref{figure:fig2}). The third most
prominent emission line detected at $\sim14982$ \AA$ $ was
identified as N$\,${\scriptsize III]}\,$\lambda$1750 redshifted
relative to Mg$\,${\scriptsize II} by $\sim3500$\,km\,s$^{-1}$. As
we explain below, this line was identified from the comparison of
the spectrum of ULAS\,J0816+2134 with those of SDSS quasars with
strong nitrogen emission lines. In particular, in the catalog of
\citet{2008ApJ...679..962J} of nitrogen-rich quasars discovered by
the SDSS survey (the quasars defined as those with the
N$\,${\scriptsize IV]}\,$\lambda$1486 and N$\,${\scriptsize
III]}\,$\lambda$1750 lines of EW$_{0} > 3$\AA), we found the
quasar SDSS\,J0852+4734 at $z_{\rm Mg\,{\scriptsize II}}=2.428$
showing the N$\,${\scriptsize III]}, C$\,${\scriptsize IV}, and
Mg$\,${\scriptsize II} emission lines of similar widths and
strengths as those of ULAS\,J0816+2134. The redshifted and scaled
spectrum of SDSS J0852+4734 is overplotted in
Figure~\ref{figure:fig1}a for comparison.

Additionally, we compared the spectrum of ULAS\,J0816+2134 with
those of quasars SDSS\,J0916+3908 and SDSS\,J1550+4238, both of
which show the prominent N$\,${\scriptsize III]} line. The
redshifts, spectral slopes, and luminosities of the comparison
quasars SDSS J0852+4734, SDSS\,J0916+3908, and SDSS\,J1550+4238
are given in Table~\ref{table:table2}. From the spectra retrieved
from the SDSS DR14 public archive \citep{2018ApJS..235...42A}, we
found that these quasars show the N$\,${\scriptsize III]} emission
line of EW$_{0}>10$ (see Figure~\ref{figure:fig3} in
Section~\ref{subsection:continuumfit}) and have absolute
magnitudes similar to that of ULAS\,J0816+2134 (see
Table~\ref{table:table2}). The detailed analysis of the spectra of
these quasars revealed that in addition to the narrow central
Gaussian-like emission component, the wings of their
N$\,${\scriptsize III]} emission lines show the presence of the
underlying broad double-peak emission component. The blueshifted
and redshifted emission peaks of the latter component are
separated by $\sim 9000$\,km\,s$^{-1}$. Furthermore, the
comparison of the SDSS spectra of SDSS\,J0852+4734 obtained at two
different epochs in 2001 and 2012 revealed significant variability
of the broad component of its N$\,${\scriptsize III]} line (see
Figure~\ref{figure:fig3}). In particular, we find that in 2001 the
strength of the blueshifted emission peak of the N$\,${\scriptsize
III]} line of SDSS\,J0852+4734 was nearly comparable to that of
ULAS\,J0816+2134 (the gray curve in Figure~\ref{figure:fig3}),
while in 2012 both blueshifted and redshifted emission peaks of
SDSS\,J0852+4734 were relatively weak (the green curve in
Figure~\ref{figure:fig3}).

The double-peak profiles of the emission lines of quasars similar
to that of the N$\,${\scriptsize III]} line of SDSS\,J0852+4734,
SDSS\,J0916+3908, and SDSS\,J1550+4238 are not unusual. Similar
line shapes show the profiles of the H$\alpha$ and H$\beta$ lines
of $\sim3$ per cent of broad line quasars \citep[e.g.,
][]{1994ApJS...90....1E,2003AJ....126.1720S,2009NewAR..53..191S,2021MNRAS.505.1029O}.
In the accretion disc wind scenario, the double-peak line profiles
are attributed to the emission from the broad line region clouds
formed in the outflow from the accretion disk and emitting in the
vicinity of the accretion disk plane
\citep[e.g.,][]{2011A&A...525L...8C,2014MNRAS.438.3340E}. The
double-peak shapes of these lines are usually well reproduced by
the emission model of the rotating relativistic disk, which
predicts the stronger blueshifted peak and the weaker redshifted
peak
\citep[e.g.,][]{1989ApJ...344..115C,1989ApJ...339..742C,1995ApJ...438..610E}.

Likewise, we attributed the non-Gaussian broad (double-peak)
emission component of the N$\,${\scriptsize III]} line of the
quasars SDSS J0852+4734, SDSS\,J0916+3908, and SDSS\,J1550+4238 to
the emission originating in the close vicinity of the accretion
disk. We modelled the N$\,${\scriptsize III]} line profiles of
these quasars both with a single Gaussian and with the model of
the circular relativistic accretion disk using the equations of
\citet{1989ApJ...344..115C}. The resulting modeled profiles are
shown in Figure~\ref{figure:fig3}
(Section~\ref{subsection:continuumfit}). The line profiles
predicted by the disk model helped us to explain the large
redshift of the N$\,${\scriptsize III]} line of ULAS\,J0816+2134
relative its Mg$\,${\scriptsize II} line. We briefly summarize the
parameters of this model. The disk model has five parameters: the
inclination angle of the disk $i$, inner and outer radii of the
emitting region $r_{\rm inner}$ and $r_{\rm outer}$ in the units
of gravitational radius $r_{\rm g}=GM_{\rm BH}/c^{2}$ (where
$M_{\rm BH}$ is the black hole mass and $G$ is the gravitational
constant), velocity dispersion $\sigma$ of the rest-frame Gaussian
line profile characterizing local turbulent broadening, and index
$q$ of the power-law surface emissivity of the disk
$\epsilon(r)\propto r^{-q}$. Following
\citet{1994ApJS...90....1E}, the value of the emissivity index was
fixed to $q = 3$ expected for local gravitational energy
dissipation in the disk \citep[see, e.g.,][]{1989ApJ...339..742C}.
The disk model predicts similar inclination angles
$i\approx40\degr$, velocity dispersions
$\sigma\approx270$\,km\,s$^{-1}$, and $r_{\rm
inner}\approx1300r_{\rm g}$ of the N$\,${\scriptsize III]}
emitting region of SDSS J0852+4734, SDSS\,J0916+3908, and
SDSS\,J1550+4238, while $r_{\rm outer}$ (which determines the
separation between the blueshifted and redshifted peaks) is
different for these quasars. We calculated $r_{\rm
outer}\approx3100$, 4600, and $2600r_{\rm g}$ for the quasars SDSS
J0852+4734, SDSS J0916+3908, and SDSS J1550+4238, respectively.
For comparison, from the analysis of the double-peak profiles of
the H$\alpha$ line of 116 SDSS quasars,
\citet{2003AJ....126.1720S} found disk inclinations in a range of
$20\degr<i<50\degr$, inner radii of $r_{\rm
inner}\sim200-800r_{\rm g}$, outer radii of $r_{\rm
outer}\gtrsim2000r_{\rm g}$, and turbulent broadening in a range
of $780-1800$\,km\,s$^{-1}$. Note that the model of the circular
rotating relativistic disk does not reproduce the central
Gaussian-like emission component of the N$\,${\scriptsize III]}
line of our comparison quasars. A similar central component is
also typically present in the double-peak profiles of the
H$\alpha$ and H$\beta$ lines of quasars and is usually fitted
separately \citep{2003AJ....126.1720S}. This central emission
component may originate in a separate region or at higher
distances from the disk plane
\citep[e.g.,][]{2014MNRAS.438.3340E}.

The N$\,${\scriptsize III]} emission line of ULAS\,J0816+2134
overlaps with the red wing of the Gaussian model and also with the
redshifted emission peak of the disk model of the
N$\,${\scriptsize III]} line of all three comparison SDSS quasars.
As seen in Figure~\ref{figure:fig3}, the velocity offset of the
N$\,${\scriptsize III]} line of ULAS\,J0816+2134 corresponds to
the typical velocity offsets of the redshifted emission peaks of
the N$\,${\scriptsize III]} line of the three comparison quasars.
We therefore conclude that first, the identification of this line
as N$\,${\scriptsize III]}\,$\lambda$1750 was most likely correct.
Second, the detected N$\,${\scriptsize III]} emission of
ULAS\,J0816+2134 seems to represent only the portion of the line.
Namely, it represents the red wing or redshifted disk emission of
the line, while the bluer part of the line is not seen, probably
due to absorption in the broad line region of the quasar. Thus, as
the bluer emission of the line is not detected, the line appears
to be significantly redshifted.

The C$\,${\scriptsize III]}\,$\lambda$1909 emission line of
ULAS\,J0816+2134 was not detected. We expect it to be broad
($7500-10000$\,km\,s$^{-1}$), similar to the C$\,${\scriptsize
III]} line of the quasars SDSS J0852+4734, SDSS\,J0916+3908, and
SDSS\,J1550+4238. Assuming that the width and flux of the
C$\,${\scriptsize III]} line of ULAS\,J0816+2134 is similar to
those of the SDSS comparison quasars, the predicted observed flux
of this line is $\sim140\times10^{-18}$\,erg\,s$^{-1}$\,cm$^{-2}$
(calculated using the average width and average rest-frame flux of
the C$\,${\scriptsize III]} line of the three comparison quasars).
This value is also comparable to the $3\sigma$ upper limit on the
flux of the C$\,${\scriptsize III]} line of ULAS\,J0816+2134
derived in Section~\ref{subsection:continuumfit} from the observed
spectrum (see Table~\ref{table:table3}). The predicted Gaussian
profile of the C$\,${\scriptsize III]} line is shown in
Figure~\ref{figure:fig1}a. As seen in this figure, the
C$\,${\scriptsize III]} line of ULAS\,J0816+2134 is likely both
too broad and weak to be detected at the signal-to-noise ratio of
the current spectrum.

\subsection{Properties of the Continuum and Emission Lines}\label{subsection:continuumfit}

We fitted the continuum of ULAS\,J0816+2134 with the power-law
model representing the UV/optical emission of the accretion disk
\citep[e.g.,][]{1982ApJ...254...22M, 1983ApJ...268..582M} with the
additional contribution from the UV Fe$\,${\scriptsize II} blended
emission lines at $\lambda_{\rm rest}=2200-3500$\,\AA:

\begin{equation}
f_{\lambda}^{\rm cont} = f_{\rm 0}^{\rm
PL}\lambda^{\alpha_{\lambda}} + f_{0}^{\rm Fe}f_{\lambda}^{\rm
Fe}, \label{eq:eq1}
\end{equation}

where $\alpha_{\rm \lambda}$ is the spectral slope, $f_{\rm
0}^{\rm PL}$ and $f_{\rm 0}^{\rm Fe}$ are the scaling factors of
the power-law component and the Fe$\,${\scriptsize II} emission,
and $f_{\lambda}^{\rm Fe}$ is the tabulated Fe$\,${\scriptsize II}
UV emission template of \citet{2006ApJ...650...57T}. The spectra
of the comparison SDSS quasars were also fitted with this model.
We performed several iterations of the fit, each time shifting the
Fe$\,${\scriptsize II} template by the redshift measured from the
Mg$\,${\scriptsize II} line fitted with a Gaussian to the
continuum-subtracted spectrum, until the best solution is reached.
The Fe$\,${\scriptsize II} template was broadened to matched the
line resolution the Mg$\,${\scriptsize II} line of
ULAS\,J0816+2134 by convolving with a Gaussian of a full width at
half maximum (FWHM) of $\sim2500$\,km\,s$^{-1}$. The contribution
of the Balmer continuum at $\lambda_{\rm rest}=2000-4000$\,\AA $ $
representing the emission of partially optically thick clouds
\citep[e.g.,][]{1985ApJ...288...94W} was neglected. From the
analysis of the spectra of the comparison SDSS quasars similar to
that of \citet{2007ApJ...669...32K} and
\citet{2011ApJ...739...56D}, we estimated that the Balmer
continuum contributes only $\lesssim10$\,per cent of the power-law
component. Additionally, note that accurate modeling of the Balmer
continuum of ULAS\,J0816+2134 and those of the comparison quasars
is difficult as their spectra do not cover the Balmer edge at
$\lambda_{\rm rest}=3646$\,\AA.

The fit of the UV/optical continuum of ULAS\,J0816+2134 in the
wavelength intervals not affected by emission lines and
atmospheric absorption ($\lambda_{\rm rest}\sim1550-1583$,
$1949-2111$, and $2323-2480$\,\AA) resulted in a slope of
$\alpha_{\lambda}=-0.63\pm0.44$. The spectral slopes of the
power-law continua of ULAS\,J0816+2134 and those of the SDSS
quasars are provided in Table~\ref{table:table2}. We find that
within the uncertainty, the continua of ULAS\,J0816+2134 and
quasar SDSS\,J0852+4734 are remarkably similar across the $H$ and
$K$ bands (see Figure~\ref{figure:fig1}a). In the $J$ band, the
spectrum of ULAS\,J0816+2134 shows the decrease in the continuum
flux at $\lambda_{\rm obs}\lesssim1.4$\,$\mu$m that may be caused
by absorption in the circumnuclear region of the quasar (see
Section~\ref{variable_absorption}). This decrease in flux
complicated the continuum fit of ULAS\,J0816+2134, while that of
SDSS\,J0852+4734 was more robust. Therefore, for the measurement
of the properties of the emission lines of ULAS\,J0816+2134, we
adopted a better measured spectral slope of SDSS\,J0852+4734 of
$\alpha_{\lambda}=-0.71$.

\begin{figure}[ht] \vspace{-5cm} \hspace{4cm}
\epsscale{2.5}\plotone{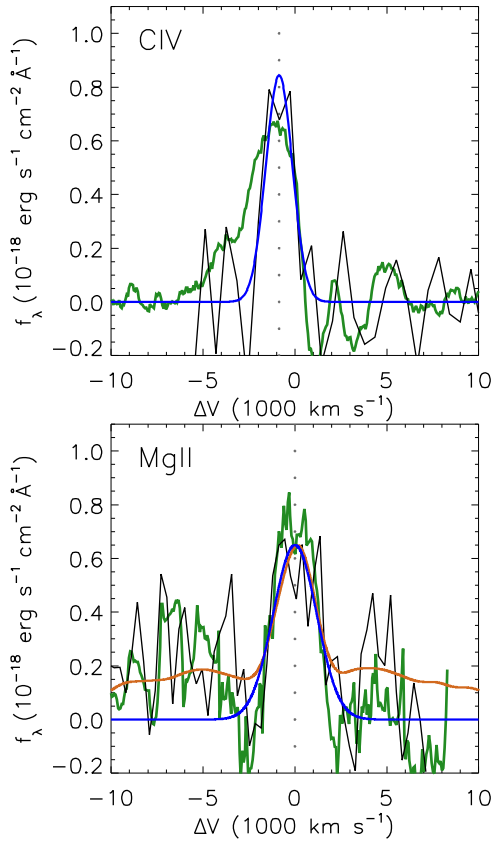} \caption{C$\,${\scriptsize IV}
and Mg$\,${\scriptsize II} line profiles of ULAS\,J0816+2134
(black lines) after subtracting the underlying power-law continuum
model. The Gaussian models of the lines are shown in blue. The
Gaussian model of the Mg$\,${\scriptsize II} line with the
contribution from the Fe$\,${\scriptsize II} UV emission is shown
with a brown line. The C$\,${\scriptsize IV} and
Mg$\,${\scriptsize II} emission lines of the low-redshift quasar
SDSS J0852+4734, are overplotted in green for comparison. For this
comparison the spectrum of SDSS J0852+4734 was redshift-corrected
and normalized to the $K$-band flux of ULAS\,J0816+2134. The
vertical dotted lines mark the velocity offset relative the
Mg$\,${\scriptsize II} line of ULAS\,J0816+2134.
\label{figure:fig2}}
\end{figure}

\begin{figure*}[ht] \vspace{-10cm} \hspace{0cm}
\epsscale{1.1}\plotone{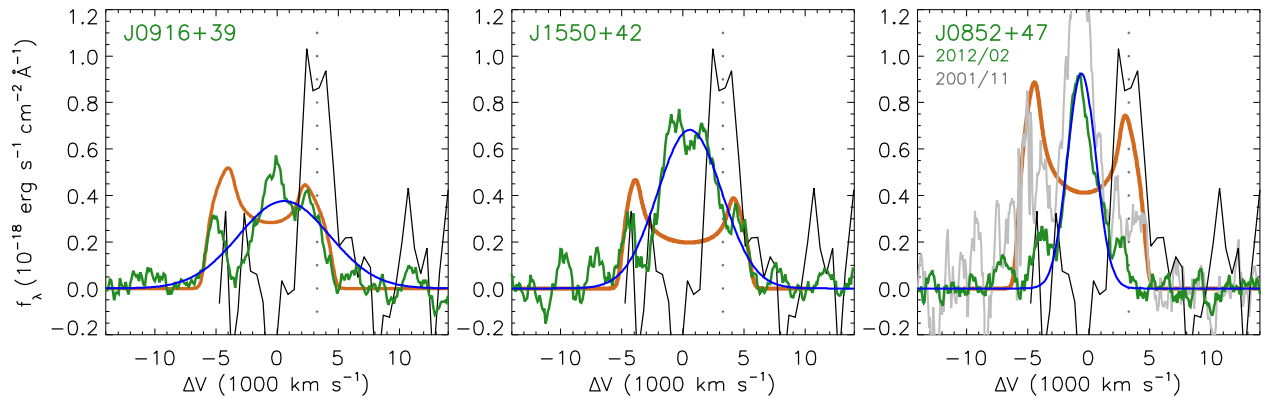} \caption{N$\,${\scriptsize III]}
line profile of ULAS\,J0816+2134 (black lines) in comparison to
those of the quasars SDSS J0916+3908 (left panel), SDSS J1550+4238
(middle panel), and SDSS J0852+4734 at epochs 2001 and 2012 (gray
and green lines in the right panel). The power-law continuum was
subtracted from the line profiles of the quasars. The Gaussian
models of the N$\,${\scriptsize III]} line of SDSS J0916+3908,
SDSS J1550+4238, and SDSS J0852+4734 (epoch 2012) are shown with
blue lines. The accretion disk models of the N$\,${\scriptsize
III]} line of these quasars are shown with brown lines. The
vertical dotted lines mark the velocity offset relative the
Mg$\,${\scriptsize II} line of ULAS\,J0816+2134.
\label{figure:fig3}}
\end{figure*}

The emission lines of ULAS\,J0816+2134 were fitted with the
Gaussian profiles after subtracting the continuum model from the
spectrum. The fitted line profiles of the C$\,${\scriptsize IV}
and Mg$\,${\scriptsize II} lines are shown in
Figure~\ref{figure:fig2}. For the Mg$\,${\scriptsize II} line we
obtained two models: the Gaussian model fitted to the power
law-subtracted spectrum and the Gaussian model with the
contribution from the Fe$\,${\scriptsize II} UV emission template
both of which are shown in Figure~\ref{figure:fig2}.  The total
flux and width of the emission lines were estimated as the
Gaussian area and FWHM, respectively. The rest-frame equivalent
widths EW$_{0}$ of the C$\,${\scriptsize IV}, N$\,${\scriptsize
III]}, and Mg$\,${\scriptsize II} lines were estimated from the
ratio between the observed flux of each line and the underlying
continuum corrected for the redshift. The properties of the
emission lines are summarized in Table~\ref{table:table2}, in
which the quoted uncertainties are the errors of the fit. For the
redshift of the ULAS\,J0816+2134 we adopted the redshift measured
from the Gaussian model of the Mg$\,${\scriptsize II} line,
$z=7.461\pm0.007$.

As discussed in Section~\ref{sec:classification}, the profile of
the N$\,${\scriptsize III]} emission line of ULAS\,J0816+2134 is
likely modified by the line-of-sight absorption and the detected
line emission may represent only the red wing of the line.
Therefore, the flux and equivalent width of the N$\,${\scriptsize
III]} line derived by us from the Gaussian fit should be
considered as the lower limits. Even based on the estimated
equivalent width of the detected portion of the N$\,${\scriptsize
III]} line (EW$_{0}\sim12.5$\,\AA; see Table 2), ULAS\,J0816+2134
can be classified as a nitrogen-rich quasar
\citep{1980ApJ...237..666O,2004AJ....128..561B,2008ApJ...679..962J}.

From the scaled Fe$\,${\scriptsize II} template, we also estimated
the Fe$\,${\scriptsize II} UV emission flux of ULAS\,J0816+2134 in
the rest-frame interval $2200-3090$\AA. This interval was adopted
for comparison with the previous measurements of the ion relative
abundance in high-redshift quasars \citep[e.g.,
][]{2007ApJ...669...32K,2011ApJ...739...56D}. Note, however, that
the Fe$\,${\scriptsize II} UV emission template was scaled using
the narrower wavelength interval ($\lambda_{\rm rest}\sim
2323-2858$\AA) than $2200-3090$\AA $  $ due to the smaller
coverage of the GNIRS spectrum and the influence of atmospheric
absorption at $\lambda_{\rm obs}\sim18000-19500$\,\AA. The
estimated Fe$\,${\scriptsize II}/Mg$\,${\scriptsize II} ratio of
ULAS\,J0816+2134 (Fe$\,${\scriptsize II}/Mg$\,${\scriptsize II}$
\sim5.7$) is in relatively good agreement with the mean value
Fe$\,${\scriptsize II}/Mg$\,${\scriptsize II} $\sim$ $3.6\pm1.3$
of the nitrogen-rich quasars from the catalog of
\citet{2008ApJ...679..962J} derived using the same
Fe$\,${\scriptsize II} UV emission template.

The $M_{1450}$ absolute magnitude of ULAS\,J0816+2134 estimated
from the flux of the $f_{\lambda}\propto \lambda^{-0.71}$
continuum at $\lambda_{\rm rest}=1450$\,\AA $  $is provided in
Table~\ref{table:table2}. The quoted error takes into account the
1$\sigma$-uncertainty of the spectrum and uncertainty in the flux
calibration of $\sim0.03$\,mag.

\begin{deluxetable}{lr@{$\pm$}lr@{$\pm$}lr@{$\pm$}l}
\tabletypesize{\footnotesize}
\tablewidth{\textwidth} \tablecaption{Total observed fluxes,
widths, and rest-frame equivalent widths of the emission lines of
ULAS\,J0816+2134. \label{table:table3}}

\tablehead{\colhead{Line}                          &
          \multicolumn{2}{c}{10$^{-18}\times$$F_{\rm line}$}                            &
           \multicolumn{2}{c}{FWHM}            &
           \multicolumn{2}{c}{EW$_{0}$}              \\
           \colhead{}                      &
           \multicolumn{2}{c}{(erg\,s$^{-1}$cm$^{-2}$)}     &
           \multicolumn{2}{c}{(km\,s$^{-1}$)}  &
           \multicolumn{2}{c}{(\AA)}                   }
\startdata
C$\,${\scriptsize IV}$\lambda$1549   & $66.4$  &  $7.4$ & $1698$ &   $275$  & $ 6.1$  & $0.7$ \\
N$\,${\scriptsize III]}$\lambda$1750 & $142.8$ &  $4.7$ & $2828$ &   $361$ & $ 12.5$ & $0.4$  \\
C$\,${\scriptsize III]}$\lambda$1909 & \multicolumn{2}{c}{$<137.0$\tablenotemark{a}} & \multicolumn{2}{c}{$\sim8700$\tablenotemark{a}} &  \multicolumn{2}{c}{$<16.2$\tablenotemark{a}}   \\
Mg$\,${\scriptsize II}$\lambda$2799\tablenotemark{b}  & $152.0$  & $14.2$  & $2784$ &   $258$ & $ 17.6$ & $1.8$ \\
Mg$\,${\scriptsize II}$\lambda$2799\tablenotemark{c}  & $118.5$  & $10.9$  & $2434$ &   $170$ & $ 14.3$ & $1.4$ \\
Fe$\,${\scriptsize II}$\lambda$$2200-3090$& $669.6$ &
$45.8$\tablenotemark{d} & \multicolumn{2}{c}{ }  & $79.7$ &$6.9$  \\
\enddata
\tablenotetext{a}{The $3\sigma$ upper limits on the flux and
equivalent width of the C$\,${\scriptsize III]} line estimated
from the spectrum. The expected line width was estimated as the
average line width of the C$\,${\scriptsize III]} line of the
comparison SDSS quasars.} \tablenotetext{b}{Based on the Gaussian
model of the Mg$\,${\scriptsize II} line.}\tablenotetext{c}{Based
on the Gaussian model of the Mg$\,${\scriptsize II} line with the
contribution from the Fe$\,${\scriptsize II} UV emission.}
\tablenotetext{d}{The Fe$\,${\scriptsize II} UV flux estimated by
integrating the flux of the fitted Fe$\,${\scriptsize II} emission
template within an interval of $\lambda_{\rm
rest}=2200-3090$\,\AA.}
\end{deluxetable}

\subsection{Variation of the Near-infrared Flux}\label{subsection:jvariation}

The brightness variations of ULAS\,J0816+2134 are seen in the
previous UKIDSS LAS observations of this quasar in $2007-2008$ and
the recent photometric data collected by us between 2021 and 2022.
Here, we focused on the analysis of the $J$, $H$, and $K$ light
curves of the $2021-2022$ observing period. These recent data
revealed the $J$-band brightness variations of ULAS\,J0816+2134 of
$\sim0.4$\,mag on timescales of a few months. During the same
$2021-2022$ epochs, no corresponding brightness variations were
detected in the $H$ and $K$ bands (see Figure~\ref{figure:fig1}b).

The variations of the UV/optical continua of quasars observed in
nearby bands are expected to produce correlated brightness changes
of comparable amplitudes. For comparison, between 2001 and 2012
the flux of the UV/optical continuum of SDSS\,J0852+4734 changed
by approximately the same amount ($\sim0.8$\,mag) at the
rest-frame wavelengths corresponding to the $J$, $H$, and $K$
bands at the redshift of ULAS J0816+2134. As such correlated
variations were not observed in the $2021-2022$ light curves of
ULAS J0816+2134, we concluded that the detected $J$-band
brightness changes were not caused by quasar variability, but most
likely have a different origin.

The observational data indicate that the variation of the
continuum flux occurs mainly at wavelengths $\lambda_{\rm
obs}\lesssim 1.4$\,$\mu$m, at which the spectrum of
ULAS\,J0816+2134 shows the signature of absorption. As seen in
Figure~\ref{figure:fig1}a, at $\lambda_{\rm obs}\lesssim
1.4$\,$\mu$m the UV continuum of ULAS\,J0816+2134 drops below the
flux level predicted by the $f_{\lambda}\propto\lambda^{-0.71}$
continuum model. Since the rest-frame effective wavelength of the
$J$ band corresponds to the region of the absorbed continuum
($\lambda_{\rm rest}\approx 1475$\AA), we attributed the observed
$J$-band brightness changes to the variation in the amount of
absorption and/or obscuration of the UV continuum emission of the
quasar with time.

In Figure~\ref{figure:fig4}, we compared the spectral shape of the
UV/optical continuum of ULAS\,J0816+2134 with those of the
comparison quasars SDSS\,J0852+4734 obtained at epochs 2001 and
2012, SDSS\,J0916+3908, and SDSS\,J1550+4238. The spectra of these
quasars were scaled to the rest-frame $K$-band flux of
ULAS\,J0816+2134. As seen in this figure, the variation of the
rest-frame $J$-band flux indicates the change of the continuum
shape of ULAS\,J0816+2134 at $\lambda_{\rm rest}\lesssim
0.17$\,$\mu$m from the heavily reddened  to the less reddened
bluer continuum similar to the continua of the comparison quasars
SDSS\,J0852+4734, SDSS\,J1550+4238 and SDSS\,J0916+3908 in the
same rest-frame wavelength interval. These variations in the
absorbed/obscured continuum flux are usually associated with the
line-of-sight variation in the gas column density caused by the
motion of the circumnuclear gas clouds of quasars. Thus, the
analysis of the multi-epoch X-ray variations of Seyfert galaxies
by \citet{2002ApJ...571..234R} revealed that the variations in the
absorbing column density are quite common among Seyferts. Several
studies have also reported the observation of the occultation of
the X-ray continuum source by individual clouds in the broad line
regions of Seyfert galaxies \citep[e.g.,
][]{2007ApJ...659L.111R,2009ApJ...696..160R,2011MNRAS.410.1027R,2021ApJ...908L..33G}.

Compared to the X-ray emitting region, the UV emitting region of
quasars is expected to be less affected by cloud
absorption/obscuration due to the larger size. Nevertheless, the
photoionization calculations of \citet{2012MNRAS.424.2255W} showed
that the UV continua of quasars can also be affected by
absorption/obscuration by the clouds of the broad line region
moving through the line of sight. Moreover, these calculations
predict similar absorption features in the obscured continua of
quasars as those seen in the rest-frame UV spectrum of
ULAS\,J0816+2134. In Section~\ref{variable_absorption}, from the
amplitude of the observed $J$-band brightness variations of the
rest-frame UV continuum, we constrained the relative variation in
the gas column density of ULAS\,J0816+2134.

\section{DISCUSSION}
\label{sec:discussion}

\subsection{Black hole mass and Eddington ratio}\label{sec:BHmass}

The method of the SMBH mass measurement of quasars from
single-epoch spectra is based on the relation between black hole
mass, widths of broad emission lines, and luminosity of the
underlying continuum
\citep{2002ApJ...571..733V,2006ApJ...641..689V}. This relation was
derived from the assumption of the virial motion of gas in the
broad line region, $M_{\rm BH}\sim~$FWHM$^{2}$~$R_{\rm BLR}
G^{-1}$ (where $R_{\rm BLR}$ is the radius of the broad line
region), and the empirical relation between continuum luminosity
and broad line region radius found from reverberation mapping of
low-redshift AGNs, $R_{\rm BLR}\propto L^{\alpha}$ \citep[where
$\alpha\approx0.5$; ][]{2005ApJ...629...61K,2006ApJ...641..689V}.
The method is commonly used for the estimation of the mass of the
SMBHs of the most distant quasars \citep[e.g.,
][]{2007ApJ...669...32K,2014ApJ...790..145D,2019ApJ...873...35S},
as their spectral properties such as the slope of the UV/optical
continuum, the width and relative strength of the broad emission
lines, are usually remarkably similar to those of low-redshift
quasars, suggesting a similar physical origin.

To estimate the mass of the SMBH of ULAS\,J0816+2134, we used the
relation of \citet{2009ApJ...699..800V} obtained for the
Mg$\,${\scriptsize II} broad emission line:

\begin{equation}\label{eq:eq2}
\frac{M_{\rm BH}}{M_{\rm \bigodot}} = 10^{6.86}\left[\frac{\rm
FWHM_{Mg\,{\scriptsize II}}}{1000~\rm km~s^{-1}}\right]^{2}
\left[\frac{\lambda L_{\lambda,3000}}{10^{44}~\rm
erg~s^{-1}}\right]^{0.5},
\end{equation}

where $\lambda L_{\lambda,\rm 3000}$ is the monochromatic
continuum luminosity at $\lambda_{\rm rest}=3000$\,\AA $ $ and
FWHM$_{\rm Mg\,{\scriptsize II}}$ is the width of the
Mg$\,${\scriptsize II} line. The $1\sigma$ scatter of this
relation is $\sim0.55$\,dex. As there is evidence that the
observed UV continuum of ULAS\,J0816+2134 is likely reddened by
the circumnuclear absorption/obscuration, the UV/optical
luminosity of this quasar measured from the observed continuum may
be uncertain. We, therefore, estimated the continuum luminosity of
ULAS\,J0816+2134, from the Mg$\,${\scriptsize II} line using the
relation between monochromatic luminosity $\lambda L_{\lambda,\rm
3000}$ and luminosity of the Mg$\,${\scriptsize II} line of
\citet{2011ApJS..194...45S}:

\begin{equation}\label{eq:eq3}
\rm log\left(\frac{\it \lambda L_{\lambda,\rm 3000}}{\rm
erg~s^{-1}}\right) = 1.22+1.016\times \rm log\left(\frac{\it
L_{\rm Mg\,{\scriptsize II}}}{\rm erg~s^{-1}}\right).
\end{equation}

The $1\sigma$ scatter of the relation is $\sim0.16$\,dex. The
bolometric luminosity of ULAS\,J0816+2134 was then calculated as
$L_{\rm bol}=5.15~\lambda L_{\lambda,\rm 3000}$ using the
bolometric correction factor of \citet{2008ApJ...680..169S}. The
Eddington ratio was computed as a ratio between the bolometric
luminosity $L_{\rm bol}$ and Eddington luminosity $L_{\rm
Edd}=1.26\times 10^{38}$($M_{\rm
BH}$/$M_{\bigodot}$)\,erg\,s$^{-1}$
\citep[][]{1997iagn.book.....P}. The estimated luminosity, black
hole mass, and Eddington ratio of ULAS\,J0816+2134 are summarized
in Table~\ref{table:table4}. This table also provides the SMBH
masses and accretion rates of the comparison SDSS quasars from
Table~\ref{table:table2}. The quoted errors were calculated based
on the uncertainties of the flux and width of the
Mg$\,${\scriptsize II} line and do not include the uncertainties
of relations~\ref{eq:eq2} and \ref{eq:eq3}.

\begin{deluxetable*}{lr@{$\pm$}lr@{$\pm$}lr@{$\pm$}lr@{$\pm$}lr@{$\pm$}l}
\tabletypesize{\footnotesize}
\tablewidth{\textwidth} \tablecaption{\label{table:table4}
Bolometric luminosity, black hole mass, and Eddington ratio of
ULAS\,J0816+2134 and the comparison SDSS quasars. }

\tablehead{\colhead{ }                          &
           \multicolumn{2}{c}{ULASJ\,0816+2134\tablenotemark{a}}           &
           \multicolumn{2}{c}{ULASJ\,0816+2134\tablenotemark{b}}            &
           \multicolumn{2}{c}{SDSS\,J0916+3908}              &
           \multicolumn{2}{c}{SDSS\,J1550+4238}            &
           \multicolumn{2}{c}{SDSS\,J0852+4734}            }
\startdata
$L_{\rm bol}$  (10$^{46}$ erg s$^{-1}$)   & 4.35&0.41  &  3.38&0.31 & 2.83&0.05 &   4.36&0.27  & 3.00&0.10 \\
$M_{\rm BH}$ (10$^{8}$ $M_{\bigodot}$)    & 5.16&1.02  &  3.48&0.51 & 3.59&0.06 &   4.58&1.25 & 2.47&0.06  \\
$L_{\rm bol}/L_{\rm Edd}$                 & 0.67&0.15  &  0.77&0.13 & 0.62&0.02 &   0.67&0.15 &  0.96&0.04\\
\enddata
\tablenotetext{a}{Based on the Gaussian model of the
Mg$\,${\scriptsize II} line.}\tablenotetext{b}{Based on the
Gaussian model of the Mg$\,${\scriptsize II} line with the
contribution from the Fe$\,${\scriptsize II} UV emission.}
\end{deluxetable*}

\subsection{Nitrogen Abundance and Black Hole Growth History}\label{evol_phase}

Both N$\,${\scriptsize III]} and C$\,${\scriptsize III]} lines are
collisionally excited lines with the comparable critical number
densities ($\sim2\times10^{10}$ and $3\times10^{9}$\,cm$^{-3}$)
and ionization potentials (29.6 and 24.4\,eV)
\citep{1978ApJ...226....1B,1999ARA&A..37..487H}. The ratio of
these lines is largely insensitive to the photoionizing conditions
such as the ionization parameter and shape of the ionizing
continuum and therefore provides a good measurement of the
relative abundance of N and C \citep[e.g.,
][]{1976ApJ...204..330S}.

The N$\,${\scriptsize III]} line detected in the spectrum of
ULAS\,J0816+2134 suggests a rather high relative abundance of
nitrogen in the broad line region of this quasar. The derived
3$\sigma$ lower limit on the N$\,${\scriptsize
III]}/C$\,${\scriptsize III]} line ratio is $\sim1$ and implies a
metallicity of $>10Z_{\bigodot}$ \citep{2002ApJ...564..592H}. Note
that the detected N$\,${\scriptsize III]} emission of
ULAS\,J0816+2134 is likely only the red wing of the total line
profile (see the discussion in Section~\ref{sec:classification}).
Therefore, the measured flux and rest-frame equivalent width of
this line should be considered as the lower limits. At the same
time, the C$\,${\scriptsize III]} line is weak. Similar to the
bluer portion of the N$\,${\scriptsize III]} line,
C$\,${\scriptsize III]} may be weakened by absorption. The
intrinsic (unaffected by obscuration and absorption)
N$\,${\scriptsize III]}/C$\,${\scriptsize III]} line ratio of
ULAS\,J0816+2134 is most likely comparable to those of the
comparison quasars from Table~\ref{table:table2}. The
N$\,${\scriptsize III]}/C$\,${\scriptsize III]} line ratio of
these quasars is similarly high as that of ULAS\,J0816+2134,
N$\,${\scriptsize III]}/C$\,${\scriptsize III]} $=0.60$, 0.98, and
0.60 for SDSS\,J0916+3908, SDSS\,J1550+4238, and SDSS\,J0852+4734,
respectively. For comparison, the N$\,${\scriptsize
III]}/C$\,${\scriptsize III]} line ratio of the composite SDSS
quasar spectrum is only EW$_{0}\sim0.44$\,\AA
\citep{2001AJ....122..549V}.

Thus, ULAS\,J0816+2134 is currently the most distant quasar
showing the relatively strong N$\,${\scriptsize III]} emission
line of EW$_{0}\gtrsim12.5$\,\AA $  $ and one of a few most
distant quasars observed at the cosmic age of 0.688\,Gyr. As
implied by the high relative abundance of nitrogen in the broad
line region of ULAS\,J0816+2134, this quasar is probably seen near
the maximum phase of the nitrogen enrichment by the AGB stars,
i.e., approximately $\sim0.25$\,Gyr after the bulk of star
formation
\citep{1991ApJ...367..126C,2000ApJ...541..660H,2022arXiv220204666J}.

The approximate starburst age of ULAS\,J0816+2134 implied by the
nitrogen abundance provides an independent way to test the history
of its black hole. If the initial burst of star formation in the
host galaxy of ULAS\,J0816+2134 occurred at a cosmic age of
$\sim0.44$\,Gyr ($z\sim10.4$), the active phase of the SMBH growth
of this quasar had lasted at most $\sim0.25$\,Gyr resulting in the
final mass of $\sim 5\times 10^{8}$M$_{\bigodot}$. Assuming that
during this previous $\sim0.25$\,Gyr, the SMBH of ULAS\,J0816+2134
was growing at a constant accretion rate of 1, the initial black
hole mass of this quasar at the time of the starburst should have
been already rather massive, $\sim3\times10^{6}$M$_{\bigodot}$
\citep[calculated assuming a radiative efficiency of
$\epsilon=0.1$; see ][]{2014ApJ...784L..38M}. This initial black
hole mass favors the scenario of massive initial seeds of the
SMBHs of quasars, similar to the other high-redshift quasars found
at $z\sim7.5$ \citep[e.g.,][]{2021ApJ...907L...1W}.

\subsection{Column Density and Size of Obscuring Clouds}\label{variable_absorption}

\begin{figure*}[ht] \vspace{-4.5cm} \hspace{-0.5cm}
\includegraphics[scale=2]{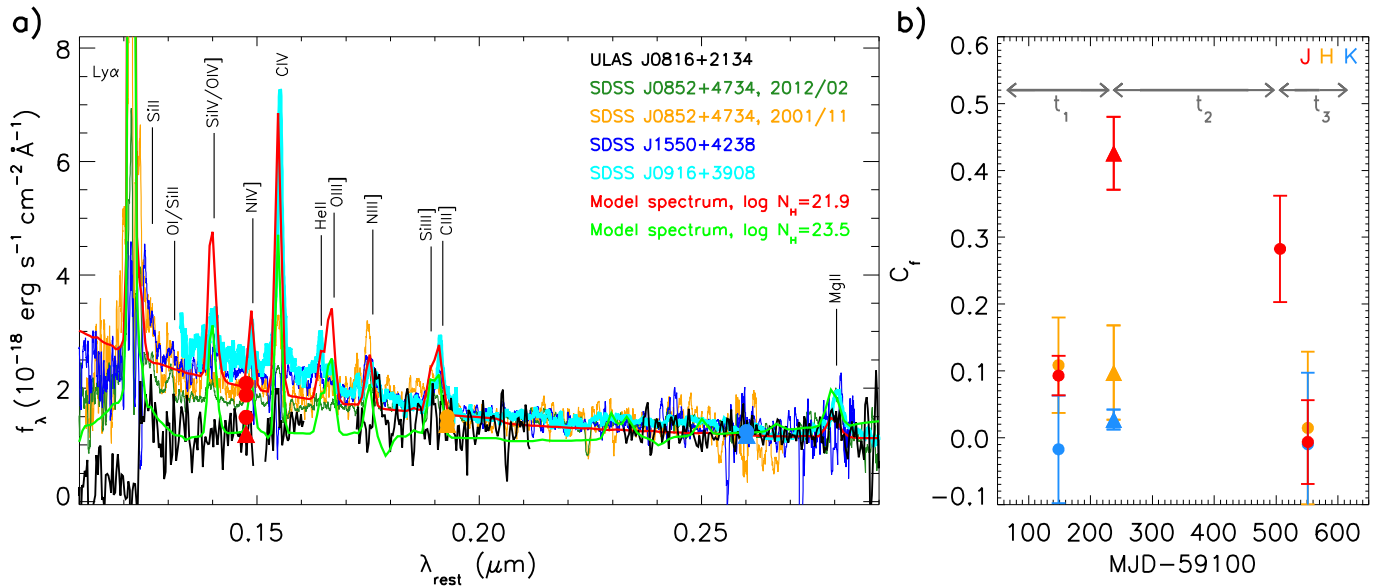}
\caption{(a) Variation of the continuum shape of ULAS\,J0816+2134
from the spectroscopic observation (black line) and $J$, $H$, and
$K$ photometry (red, orange, and blue circles, respectively). The
$J$, $H$, and $K$ fluxes measured from the spectrum of
ULAS\,J0816+2134 are depicted with a red, orange, and blue
triangle. The observed variation in the continuum flux of
ULAS\,J0816+2134 is shown in comparison with the spectra of the
quasars SDSS\,J0852+4734 obtained in 2001 and 2012 (orange and
green lines), the spectra of SDSS\,J1550+4238 (blue line), and
SDSS\,J0916+3908 (cyan line) normalized to the $K$-band continuum
flux of ULAS\,J0816+2134. The spectra predicted by the
photoionization models calculated for a column density of the
broad line region gas of $N_{\rm H}=10^{21.9}$ and
$10^{23.5}$\,cm$^{-2}$ are indicated with a red and light green
line. (b) Variation of the covering fraction of the continuum
emitting region of ULAS\,J0816+2134 in the $J$, $H$, and $K$ bands
during the occultation by an intervening cloud of the broad line
region. The circles and triangles correspond to the covering
fraction estimated from the $J$-, $H$-, and $K$-band photometry
and from the spectrum of ULAS\,J0816+2134, respectively. The
occultation time is divided into time intervals $t_{1}$, $t_{2}$,
and $t_{3}$ approximately corresponding to the ingress phase, the
phase of maximum coverage, and the egress phase of the
occultation.} \label{figure:fig4}
\end{figure*}

The amplitude of the flux variation and the duration of the
partial obscuration caused by the motion of the individual
absorbing/obscuring clouds of the broad line region through the
line of sight of the continuum source provide strong constraints
on the column density and linear size of the clouds,
e.g.,\citep{2007ApJ...659L.111R, 2009ApJ...696..160R,
2011MNRAS.410.1027R, 2021ApJ...908L..33G}. As discussed in
Section~\ref{subsection:jvariation}, the variation of the $J$-band
flux of ULAS\,J0816+2134 can be attributed to the variation in the
absorbing column density caused by the motion of such a cloud. To
constrain the column density of the intervening cloud from the
observed amplitude of the $J$-band flux variations, we simulated
the model spectra of ULAS\,J0816+2134 using version 13.03 of the
photoionization code CLOUDY \citep{1998PASP..110..761F}. For the
photoionizing continuum, we employed the standard AGN continuum
described in \citet{1987ApJ...323..456M}. We then adopted an
ionization parameter of $U=10^{-1.2}$, an average hydrogen number
density of the broad line region gas of $n_{\rm
H}=10^{10}$\,cm$^{-3}$, and a broad line region radius of $R_{\rm
BLR}=10^{17}$\,cm. The ionization parameter and gas number density
were estimated from the average C$\,${\scriptsize
III]}/C$\,${\scriptsize IV} line ratio of the comparison quasars
SDSS\,J0852+4734, SDSS\,J1550+4238, and SDSS\,J0916+3908, assuming
that the photoionization conditions in the broad line region of
these quasars are similar to those of ULAS\,J0816+2134. For the
chemical composition of the broad line region gas clouds, we
adopted the metallicity equal to $5Z_{\bigodot}$ \citep[typical
for the high-redshift quasars, e.g., ][]{2003A&A...398..891D,
2003ApJ...589..722D} and the metal abundance model of
\citet{1993ApJ...418...11H} for an evolving starburst. The
hydrogen column density varied in a range of $N_{\rm H} =
10^{21}-10^{24}$\,cm$^{-2}$.

The approximate radius of the broad line region used in the
calculations was derived from the typical size of the
N$\,${\scriptsize III]} emitting region of the comparison quasars.
It was estimated for an SMBH of $\sim 5\times10^{8}M_{\bigodot}$
as $R_{\rm BLR, NIII]}\sim 2000r_{\rm g}\sim1.5\times10^{17}$\,cm
(see Section~\ref{sec:classification}). The relation of
\citet{2009ApJ...699..800V} for the size of the Mg$\,${\scriptsize
II} emitting region provides the comparable estimate, $R_{\rm BLR,
MgII}\approx 8.8\times10^{17}$\,cm.

The calculated model spectra were broadened to match the line
width of the comparison SDSS quasars and compared with the
spectrum of ULAS\,J0816+2134 and its $J$-band flux at different
epochs. The model spectra predict the C$\,${\scriptsize IV},
He$\,${\scriptsize II}, N$\,${\scriptsize III]}, and
C$\,${\scriptsize III]} emission lines of similar strengths to
those of the comparison SDSS quasars. Note however that the model
overpredicts the strengths of the Si$\,${\scriptsize IV},
O$\,${\scriptsize IV]}, O$\,${\scriptsize III]}, and
Si$\,${\scriptsize III]} emission lines. The latter group of the
emission lines may originate at different physical conditions as
those assumed by us for the C$\,${\scriptsize IV},
N$\,${\scriptsize III]}, and C$\,${\scriptsize III]} emission
lines. Here, we only consider the effect of the varying column
density on the spectral shape of the UV continuum for the adopted
physical conditions.

From the model spectra, we found that the lowest $J$-band
continuum flux corresponds to a column density of $N_{\rm
H}\approx10^{21.9}$\,cm$^{-2}$ (see Figure~\ref{figure:fig4}a).
This model reproduces well the shape of the Gemini/GNIRS spectrum
of  ULAS\,J0816+2134, especially at $\lambda_{\rm rest}
\lesssim0.17$\,$\mu$m, mostly affected by absorption. Note that
the Gemini/GNIRS spectrum was obtained during the lowest $J$-band
continuum flux reduced by line-of-sight absorption/obscuration.
The highest $J$-band continuum flux measured from photometry
corresponds to a line-of-sight gas column density of $N_{\rm
H}\approx10^{23.5}$\,cm$^{-2}$. Thus, the total variation of the
gas column density associated with the observed variation in the
$J$-band flux is $\Delta N_{\rm H}\approx10^{23.49}$\,cm$^{-2}$.
The shape of the $J$-band light curve implies that the actual
amplitude of the variation in the column density could have been
higher than that provided by the data points. Therefore, we
adopted the derived relative column density variation as the lower
limit on the column density of the intervening cloud that
absorbed/obscured the UV continuum of ULAS\,J0816+2134. The
estimated value is comparable to the column density of the broad
line region clouds inferred from the X-ray observations of Seyfert
galaxies \citep[e.g.,
][]{2007ApJ...659L.111R,2009ApJ...696..160R,2011MNRAS.410.1027R}.

Assuming that the unobscured spectrum of ULAS\,J0816+2134 is
similar to that of the blue simulated spectrum corresponding to
$N_{\rm H}\approx10^{21.9}$\,cm$^{-2}$,  we estimated the
approximate fraction of the obscured $J$-band flux of the
rest-frame UV continuum of ULAS\,J0816+2134 relative to the flux
of the unobscured UV continuum for each of the observational
epochs as

\begin{equation}\label{covering_fr}
C_{\rm f}(t) = \left(1-\frac{f_{J}^{\rm
J0816}(t)}{f_{J}^{0}}\right),
\end{equation}

where $f_{J}^{\rm J0816}(t)$ and $f_{J}^{0}$ are the $J$-band flux
of ULAS\,J0816+2134 at epoch $t$ and that of the unobscured
continuum model. The variation of the estimated covering fraction
of the UV continuum caused by the motion of the intervening cloud
through the line of sight of the quasar is shown in
Figure~\ref{figure:fig4}b. As seen in the figure, the evolution of
the covering fraction was relatively smooth. It increased between
the first two epochs (February and May 2021) and then decreased
between the last two epochs (February and March 2022). The
timescale of the whole variation was$\sim 403$\,d, corresponding
to $\sim48$d in the rest-frame of the quasar. The smooth and
symmetric shape of the $J$-band variation in the covering fraction
is similar to the covering fraction variation of the X-ray
continuum emitting region observed by \citet[e.g.,
][]{2007ApJ...659L.111R, 2021ApJ...908L..33G}. Following these
previous works, we divided the occultation time into time
intervals $t_{1}$, $t_{2}$, and $t_{3}$, approximately
corresponding to the ingress phase, the phase of maximum coverage,
and the egress phase of the occultation \citep[see Figure 2 of
][]{2007ApJ...659L.111R}. The ingress/egress time provides the
estimate of the linear cloud size. From the average of the ingress
and egress rest-frame time ($t_{1,3}=0.5\times
(t_{1}+t_{3})\approx 7.9)$\,d, we estimated a cloud size of
$D_{\rm cl}=t_{1,3}\upsilon_{\rm K}=1.7\times 10^{14}$\,cm, where
$\upsilon_{\rm K}$ is the Keplerian velocity of the cloud. We
assumed $\upsilon_{\rm K}\approx 2500$\,km\,s$^{-1}$, comparable
to the width of the Mg$\,${\scriptsize II} line of
ULAS\,J0816+2134. The size of the UV emitting region corresponding
to the rest-frame $J$-band wavelength ($\lambda_{\rm
rest}\approx1475$\,\AA) is then $D_{1475}\approx D_{\rm cl}\times
t_{\rm 2}/t_{1,3}\approx 6.9\times 10^{14}$\,cm, where $t_{\rm
2}\approx32$\,d in the quasar rest-frame.

We compared the radius of the continuum source at $\lambda_{\rm
rest}\approx1475$\,\AA $  $ derived from the duration of the
occultation by the intervening cloud with the half-light radius
predicted by the temperature profile of the standard accretion
disk of \citet{1973AA....24..337S} for the same emitting region.
Using equation~2 of \citet[][]{2011ApJ...729...34B}, the estimated
half-light radius of the emitting region of ULAS\,J0816+2134 at
$\lambda_{\rm rest}\approx1475$\,\AA $ $ is $R_{\rm 1475}\approx
1.6\times 10^{15}$\,cm (assuming a radiative efficiency of
$\epsilon=0.1$), which looks comparable within the uncertainties
to the size of the continuum emitting region from the occultation
light curve. Thus, our assumption about the variation in the
covering fraction due to one nearly complete occultation of the UV
continuum source of ULAS\,J0816+2134 detected in the $J$ band is
most likely correct despite the relatively poor sampling of the
occultation light curve.

\section{SUMMARY}
\label{sec:summary}

We presented the near-infrared spectroscopic and imaging
observations, and the data analysis of the newly discovered quasar
ULAS J0816+2134 at $z_{\rm MgII}=7.461\pm0.007$. Based on the
detection of the N$\,${\scriptsize III]} emission line of
EW$_{0}\gtrsim12.5$\,\AA, ULAS\,J0816+2134 was classified as a
nitrogen-rich quasar. Being the most distant nitrogen-rich quasar
and one of the most distant quasars known today, ULAS J0816+2134
provides new independent constraints on the starburst age of the
host galaxies and on the growth history of the SMBHs of
high-redshift quasars. We summarize our main results and
conclusions as follows:

1. ULAS J0816+2134 is the least luminous of the known $z\sim7.5$
quasars of an absolute magnitude of $M_{1450}=-25.33\pm0.07$. The
mass of the SMBH of ULAS J0816+2134 estimated using the
Mg$\,${\scriptsize II} emission line is $M_{\rm
BH}=(5.16\pm1.02)\times10^{8}$$M_{\bigodot}$ and the Eddington
accretion rate is $L_{\rm bol}/L_{\rm Edd}=0.67\pm0.15$.

2. The high nitrogen abundance of the broad line region of
ULAS\,J0816+2134 indicates that this quasar is probably seen near
the maximum of the nitrogen enrichment of the circumnuclear region
by the asymptotic giant branch stars, which implies a starburst
age of the host galaxy of $\sim0.25$\,Gyr. The starburst age
provides an independent constraint on the growth timescale and
initial mass of the SMBH of ULAS\,J0816+2134. It implies a
sufficiently massive initial black hole mass of $\sim3\times
10^{6}$M$_{\bigodot}$. The appearance of such massive initial
black holes in the early Universe is yet to be understood.

3. The flux variation of the observed $J$-band continuum and the
lack of the corresponding variations in the $H$ and $K$ bands,
were attributed to the variation in the line-of-sight column
density caused by the motion of an intervening broad line region
cloud.  The amplitude of the observed $J$-band flux variation
implies a cloud column density of $N_{\rm
H}\gtrsim10^{23.49}$\,cm$^{-2}$, while the duration of the
variation implies a cloud size of $D_{\rm
cl}\approx1.7\times10^{14}$\,cm. The derived column density of the
intervening cloud is consistent with the typical column density of
mostly neutral, gravitationally bound clouds of the broad line
region of quasars that may fuel the growth of their SMBHs.


\acknowledgments

This work was supported by the Ministry of Science and Technology
of Taiwan, grant Nos MOST 109-2112-M-008-021-MY3 and MOST
110-2112-M-008-021-MY3. Based on data obtained at the
international Gemini Observatory, a program of NSF's NOIRLab, via
the time exchange program between Gemini and the Subaru Telescope
(Program ID: GN-2021A-FT-106). The Subaru Telescope is operated by
the National Astronomical Observatory of Japan. The international
Gemini Observatory at NOIRLab is managed by the Association of
Universities for Research in Astronomy (AURA) under a cooperative
agreement with the National Science Foundation on behalf of the
Gemini Observatory partnership: the National Science Foundation
(United States), National Research Council (Canada), Agencia
Nacional de Investigaci\'{o}n y Desarrollo (Chile), Ministerio de
Ciencia, Tecnolog\'{i}a e Innovaci\'{o}n (Argentina),
Minist\'{e}rio da Ci\^{e}ncia, Tecnologia, Inova\c{c}\~{o}es e
Comunica\c{c}\~{o}es (Brazil), and Korea Astronomy and Space
Science Institute (Republic of Korea). \\Based on observations
obtained with UKIRT (Program IDs: U/20B/NCU02, U/21B/NCU01,
U/22A/NCU01). UKIRT is owned by the University of Hawaii (UH) and
operated by the UH Institute for Astronomy; operations are enabled
through the cooperation of the East Asian Observatory.



\end{document}